\documentclass[12pt]{article}

\usepackage{graphics}
\usepackage{graphicx}
\usepackage{amssymb}
\usepackage{epsf}
 \usepackage{setspace}  

\newcommand{\halmos}{\rule{1ex}{1.4ex}}
\newcommand{\qed}{\hfill \halmos} 
\newcommand{\text}[1]{\hbox{\rm \ #1\ \/}}
\newcommand{\be}[1]{\begin{equation}\label{#1}}
\newcommand{\ee}{\end{equation}}
\newcommand{\bi}{\begin{itemize}}
\newcommand{\ei}{\end{itemize}}
\newcommand{\ben}{\begin{enumerate}}
\newcommand{\een}{\end{enumerate}}

\newcommand{\R}{{\mathbb R}}  

\newcommand{\bl}[1]{\begin{Lemma}\label{#1}}
\newcommand{\el}{\qed\end{Lemma}}
\newcommand{\bt}[1]{\begin{Theorem}\label{#1}}
\newcommand{\et}{\end{Theorem}}
\newcommand{\epr}{\end{proof}}
\newcommand{\bpr}{\begin{proof}}
\newenvironment{proof}{\noindent {\em Proof}.\ }{\hspace*{\fill}$\halmos$\medskip}

\newcommand{\beqn}{\begin{eqnarray*}}
\newcommand{\eeqn}{\end{eqnarray*}}

\newtheorem{Theorem}{Theorem}
\newtheorem{Proposition}{Proposition}
\newtheorem{Lemma}{Lemma}
\newtheorem{Corollary}{Corollary}

\newcommand{\norma}[1]{\ensuremath{\left| #1 \right|}}

\newcommand{\C}{\mathbb{C}}

\begin{document}

\title{A dichotomy for a class of cyclic delay systems}

\author{Germ\'{a}n Andr\'{e}s Enciso\footnote{Mathematical Biosciences Institute, 231 W 18th Ave, Columbus OH 43210.  email:  genciso@mbi.osu.edu.}}

\maketitle

\paragraph{Abstract:}
Two complementary analyses of a cyclic negative feedback system with
delay are considered in this paper.  The first analysis applies the
work by Sontag, Angeli, Enciso and others regarding monotone control
systems under negative feedback, and it implies the global
attractiveness towards an equilibrium for arbitrary delays.  The
second one concerns the existence of a Hopf bifurcation on the delay
parameter, and it implies the existence of nonconstant periodic
solutions for special delay values.  A key idea is the use of the
Schwarzian derivative, and its application for the study of
Michaelis-Menten nonlinearities.  The positive feedback case is also
addressed.

\vspace{2ex}


\vspace{2ex}

Consider the cyclic nonlinear system

\be{nonlinear general}
\begin{array}{l}
\dot{x}_i= g_i(x_{i+1}) - \mu_i x_i,\  i=1\ldots n-1, \\
\dot{x}_n= g_n(x_1(t-\tau)) - \mu_n x_n,  \\
\end{array}
\ee
for $n\geq 1$. Assume that each function $g_i$ is either
increasing or decreasing and that the system is subject to negative
feedback. More formally, let

\be{assumptions} \mu_i>0,\ \delta_i g_i'(x)\geq0,\  \delta_i\in
\{1,-1\}, \ i=1\ldots n, \mbox{ and } \delta_1\cdot\ldots \cdot
 \delta_n=-1.
\ee

This system can be considered a generalization of classical models,
by Goldbeter \cite{Goldbeter:1996} and Goodwin \cite{Goodwin:1965},
of autoregulated biochemical networks under negative feedback. Delay
systems with this general structure can also be found in the
modeling of neural networks, for instance in
\cite{Pakdaman:1997,Townley:2000}, using $g_i(x)=\alpha_i
\tanh(\beta_i x)$ as nonlinearities.  It should also be noted that
different delays can be introduced in the nonlinear terms of each
equation without loss of generality, since all but one of them can
be removed with a simple change of variables.

An important special case in biochemical models is that in which
those functions $g_i(x)$ which are not linear have the
Michaelis-Menten form

\be{MM} f(x)=\frac{ax^m}{b+x^m}+c, \ \mbox{ or } \
f(x)=\frac{a}{b+x^m}+c,\ a,b>0,\ c\geq0, m=1,2,\ldots. \ee A recent
(though undelayed) model within this framework is that of the
so-called repressilator, see Elowitz and Leibler
\cite{Elowitz:Leibler:2000}. We will give especial attention below
to this type of nonlinearity.

The dynamics of the bounded solutions of system (\ref{nonlinear
general}) under assumptions (\ref{assumptions}) is governed by a
Poincare-Bendixson result, proved by Mallet-Paret and Sell in
1996~\cite{MP:Sell:JDE1996}. Informally speaking, for every initial
condition the solution of the system approaches either an
equilibrium, a periodic orbit, or a homoclinic chain of orbits. In
particular, any chaotic behavior is ruled out.  In the positive
feedback case $\delta_1\cdot \ldots \cdot \delta_n=1$, system
(\ref{nonlinear general}) is monotone and also falls within the
framework of Mallet-Paret and Sell.  A large number of results are
known in that case, the most important one perhaps being that the
generic solution is convergent towards an equilibrium
\cite{Hirsch:1988,Smith:monotone}.

The work of Sontag and Angeli \cite{Sontag:mono} can be used to
establish a relationship between the system (\ref{nonlinear
general}) and the one-dimensional discrete system

\be{discrete system} u_{k+1}=g(u_k),
 \ee
where

\be{g def} g(u):=\frac{1}{\mu_1}g_1\circ\frac{1}{\mu_2}g_2\circ
\ldots \circ \frac{1}{\mu_n}g_n. \ee
Namely, if the discrete system
(\ref{discrete system}) is globally attractive towards its unique
equilibrium $x_0$, then the original system (\ref{nonlinear
general}) is globally attractive towards its unique equilibrium, for
all values of the delay $\tau$; see also
\cite{Enciso:Sontag:DCDS2006,Enciso:Smith:Sontag:JDE2005,Enciso:Sontag:JMB2004,Angeli:Sontag:interconnections,Sontag:embedding},
and Hale and Ivanov \cite{Hale:Ivanov:1993}.

A second branch of study for systems analogous to (\ref{nonlinear
general}) is the search for nonconstant periodic oscillations. This
usually involves a different kind of assumption, namely that the
system (\ref{nonlinear general}) is `ejective' around its unique
equilibrium for large enough delay.   Such arguments usually require
the hypothesis $\norma{g'(x_0)}>1$, which in particular rules out
the global attractiveness of (\ref{discrete system}). See
Nussbaum~\cite{Nussbaum:1986}, Hadeler and
Tomiuk~\cite{Hadeler:Tomiuk:1977}, Hale and Ivanov
\cite{Hale:Ivanov:1993}, and Ivanov and Lani-Wayda
\cite{Ivanov:Lani-Wayda:2004}, among others.

In the present paper, both approaches are unified to give a more
complete picture of the relationship between system (\ref{nonlinear
general}) (under assumptions (\ref{assumptions}))  and system
(\ref{discrete system}).  A Hopf bifurcation approach is considered
to prove that $\norma{g'(x_0)}>1$ implies the existence of periodic
solutions of (\ref{nonlinear general}) for certain values of $\tau$.
Also, an important class of nonlinearities $g_i$ is shown to be such
that the following conditions are a dichotomy:
\begin{enumerate}
\item  The system (\ref{discrete system}) is globally attractive towards
$x_0$.
\item  $\norma{g'(x_0)}>1$.
\end{enumerate}
The main result of this paper is the following theorem, where $Sg$
denotes the Schwarzian derivative of the function $g$; see for
instance \cite{Sedaghat:2003}.

\begin{Theorem} \label{main theorem}
Consider a system (\ref{nonlinear general}) under assumptions
(\ref{assumptions}), and let the function $g(x)$ defined by (\ref{g
def}) be bounded.  Suppose that all nonlinear functions $g_i(x)$ are
of Michaelis-Menten form (\ref{MM}), $m\geq 1$, or that $Sg<0$. Then
exactly one of the following holds:

\begin{enumerate}
\item System (\ref{discrete system}) is globally attractive to a
unique equilibrium, and (\ref{nonlinear general}) is also globally
attractive to a unique equilibrium, for all values of the delay
$\tau$.

\item System (\ref{discrete system}) contains nonconstant
periodic solutions, and system (\ref{nonlinear general}) is subject
to a Hopf bifurcation on the delay parameter $\tau$.  In particular,
(\ref{nonlinear general}) contains nonconstant periodic solutions
for some values of $\tau$.
\end{enumerate}
\end{Theorem}

\begin{figure}
\center{\includegraphics[width=6in]{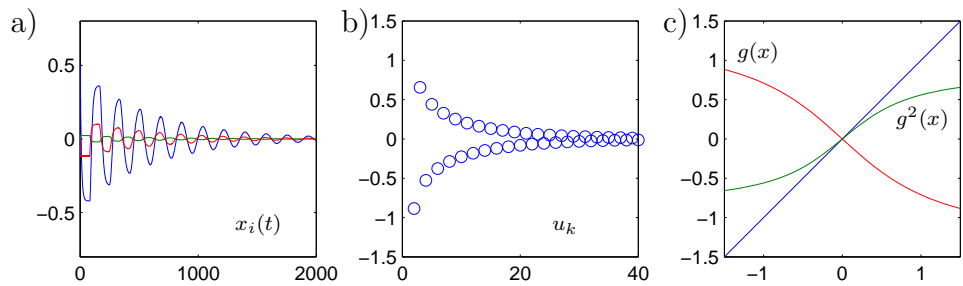}}
\begin{picture}(0,0)

\put(-165,210){a)}

\put(-40,210){b)}

\put(82,210){c)}

\put(-80,135){$\scriptstyle x_i(t)$}

\put(40,135){$\scriptstyle u_k$}

\put(110,200){$\scriptstyle g(x)$}

\put(170,175){$\scriptstyle g^2(x)$}
\end{picture}
\caption{  Typical solutions of a) system (\ref{nonlinear general})
and b) system (\ref{discrete system}), where $n=3$,
$g_1=g_2=g_3=tan^{-1}(x)$, $\mu_1=0.11$, $\mu_2=2.5$, $\mu_3=4$, and
$\tau=80$. c) The induced decreasing function $g(x)$ and the
increasing function $g^2(x)=g(g(x))$ (see Lemma~\ref{lemma
equivalence}). Here $\norma{g'(x_0)}=1/1.1<1$.}
 \label{figure 1}
\end{figure}

\begin{figure}
\label{figure 2}
 \center{\includegraphics[width=6in]{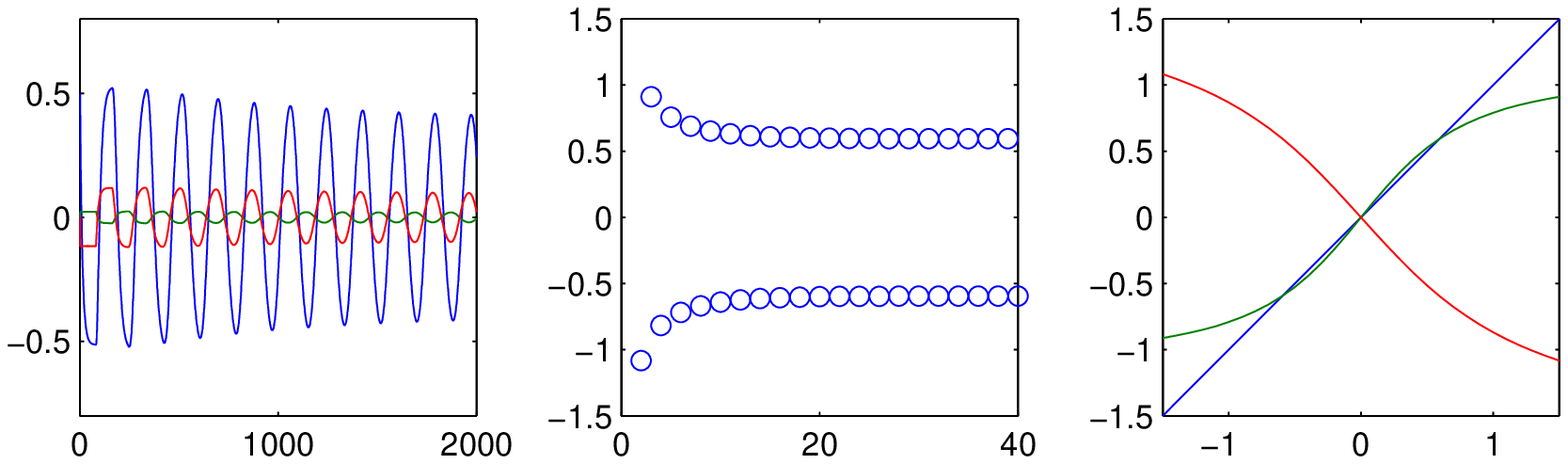}}
\begin{picture}(0,0)

\put(-165,210){a)}

\put(-40,210){b)}

\put(82,210){c)}

\put(-80,135){$\scriptstyle x_i(t)$}

\put(40,135){$\scriptstyle u_k$}

\put(110,185){$\scriptstyle g(x)$}

\put(170,175){$\scriptstyle g^2(x)$}
\end{picture}
\caption{The same system is considered as in Figure~\ref{figure 1},
except that the value of $\mu_1$ has been changed to $0.09$. The
typical solutions of a) system (\ref{nonlinear general}) and b)
system (\ref{discrete system}) now appear to be limit oscillations
and periodic 2-cycles. c) In this case $\norma{g'(x_0)}=1/0.9>1$ and
$g^2(x)=x$ has several solutions.}
\end{figure}

The Hopf bifurcation in the second case is not shown to be
supercritical, although this seems to be the case from numerical
simulations.   Bounded decreasing functions with negative Schwarzian
derivative include $-\tan^{-1}(x)$, $-\tanh(x)$, and $e^{-x}$ (on
$\R^+$), as well as their decreasing compositions. They also include
the Michaelis-Menten functions above for $m>1$, as is shown here in
Lemma~\ref{lemma MM}.

It is important to note that this information is not provided
\emph{a priori} by the Poincare-Bendixson theorem itself, which
doesn't give conditions for the different possible outcomes.  Even
knowing that the equilibrium of (\ref{nonlinear general}) is
unstable doesn't guarantee the existence of periodic oscillations,
since for instance homoclinic orbits need to be ruled out (possibly
using Morse decomposition theory \cite{Mallet-Paret:Morse}).

A corresponding theorem will be stated and proved in the positive
feedback case $\delta_1 \cdot \ldots \cdot \delta_n=1$, simplifying
a result from Angeli and Sontag \cite{Sontag:mono} in the case
without delays, and from Enciso and Sontag
\cite{Enciso:Sontag:setchars} in the delay case.

System (\ref{discrete system}) is one-dimensional and doesn't
contain delays, which makes it much more tractable than
(\ref{nonlinear general}). The assumption of the negative Schwarzian
is a common simplifying hypothesis in the discrete systems
literature; see for instance \cite{Liz:2003} for an application to
continuous systems.
 A Hopf bifurcation approach has
also been proposed in the Poincare-Bendixson context in
\cite{Wang:2005}.

The direct contributions of the present paper are i) to show that
for an important class of nonlinearities the two alternative cases
form a dichotomy; ii) to formally establish a relationship between
the discrete and the continuous system, which has already been
conjectured by Smith \cite{Smith:1987} in the undelayed case; iii)
to carry out a direct Hopf bifurcation analysis of the linear system
associated to (\ref{nonlinear general}) (which is new to my
knowledge),  and iv) to illustrate the usefulness of the Schwarzian
derivative in the context of Michaelis-Menten functions.

In Section~\ref{section Schwarzian} the concept of the Schwarzian
derivative is briefly introduced and applied to Michaelis-Menten
functions. In Section~\ref{section discrete}, the discrete system
and its relationship with (\ref{nonlinear general}) are described.
In Section~\ref{section Hopf} the Hopf bifurcation argument is
developed, and in Section~\ref{section positive} the positive
feedback case is considered.   Finally, in Section~\ref{section
future}, the relationship with the general results in
\cite{Sontag:mono} and \cite{Enciso:Sontag:DCDS2006} is shortly
discussed, and a conjecture is described from numerical simulations.

\section{$Sg$ and Michaelis-Menten Functions} \label{section Schwarzian}

An important concept related to the stability of discrete dynamical
systems is the so-called \emph{Schwarzian derivative} $Sf$ of a real
function $f$, defined by

\[Sf(x)=\frac{f'''(x)}{f'(x)} -
\frac{3}{2}\left(\frac{f''(x)}{f'(x)}\right)^2.\]

The properties of $Sf$ that will be useful here are summarized in
the following lemma;  see \cite{Sedaghat:2003}, Section~2B for
proofs and details.  Intuitively, the condition $Sf<0$ restricts the
form of the function $f$ so that the dynamics of $u_{k+1}=f(u_k)$ is
more easily determined.

\begin{Lemma} \label{lemma Sg}
Let $f,g$ be $C^3$ real functions on a real interval.  Then the
following hold:
\begin{enumerate}
\item If $Sf<0$, then $f'$ cannot have positive local
minima or negative local maxima.

\item $S(f\circ g)(x)=Sf(g(x))g'(x)^2 + Sg(x)$.

\item $Sf<0,\ Sg<0$ imply $S(f\circ g) <0$.
\end{enumerate}
\end{Lemma}

It is now shown that the class of functions with negative Schwarzian
derivative includes the cooperative Michaelis-Menten functions with
$m>1$, and that $S(x/(b+x))=0$.

\begin{Lemma}  \label{lemma MM}
Let $a,b>0, \ c\geq0, $ and $m=1,2,\ldots$, and define
\[
f(x)=\frac{a x^m}{b+x^m}+c,\ g(x)=\frac{a}{b+x^m}+c.
\]
Then $\displaystyle Sf(x)=Sg(x)=-\frac{m^2-1}{2}\frac{1}{x^2}$.
\end{Lemma}

\bpr Noting that the Schwarzian derivative doesn't change after
multiplication by or addition of a constant, we can assume that
$a=1,\ c=0$. Using the quotient rule we compute
\[
f'(x)=\frac{mx^{m-1}}{b+x^m} - \frac{m
x^{2m-1}}{(b+x^m)^2}=\frac{m}{x}(y-y^2)=\frac{m}{x}y(1-y),
\]
where $y=f(x)$.  Similarly we compute
\[
\begin{array}{l} \displaystyle
f''(x)=
       -\frac{m}{x^2}y(y-1)(2my-(m-1))
       \\ \displaystyle
f'''(x)=\frac{m}{x^3}y(1-y)[6m^2y^2+(6m-6m^2)y+(m-1)(m-2)].
\end{array}
\]

We calculate the Schwarzian derivative
\[
\begin{array}{l} \displaystyle
Sf(x)=\frac{f'''(x)}{f'(x)}-\frac{3}{2}\left(\frac{f''(x)}{f'(x)}\right)^2
\\ \displaystyle
=\frac{1}{x^2}[6m^2y^2-6m(m-1)y+(m-1)(m-2)] \\ \displaystyle
\ \ \ -\frac{3}{2}\frac{1}{x^2}[4m^2y^2 - 4m(m-1)y + (m-1)^2] \\
\displaystyle =\frac{1}{x^2}[(m-1)(m-2)-\frac{3}{2}(m-1)^2] =
-\frac{m^2-1}{2}\frac{1}{x^2}.
\end{array}
\]

To compute $Sg(x)$, it is easy to see that $g=b^{-1} f\circ \kappa$,
where $\kappa(x)=b^{1/m}/x$. A simple computation shows that
$S\kappa=S(1/x)=0$, $x\not=0$. Therefore
\[
Sg(x)=S(f\circ \kappa)=Sf(\kappa(x)) \kappa'(x)^2 + S\kappa(x) =
-\frac{m^2-1}{2}x^2\frac{1}{x^4}+0 = -\frac{m^2-1}{2}\frac{1}{x^2}.
\]
\epr

\section{The Discrete System} \label{section discrete}

Consider a continuous, bounded, decreasing function $g:I\to I$,
where $I=\R$ or $I=[a,\infty)$, $a\in \R$.  It can be easily seen
that there is a unique fixed point $x_0$ of $g$. The study of the
discrete system (\ref{discrete system}) becomes straightforward by
relating its dynamics to that of the system $u_{k+1}=g^2(u_k)$,
since the function $g^2(x)=g(g(x))$ is bounded and increasing.   We
state the following lemma for convenience; see also Angeli and de
Leenheer \cite{predator:prey} for an extended discussion.

\begin{Lemma} \label{lemma equivalence}
  System (\ref{discrete system}) is globally attractive if and only if the equation $g(g(x))=x$ has the unique solution $x_0$.
\end{Lemma}
\bpr All solutions of the system $u_{k+1}=g(g(u_k))$ are monotonic
increasing or decreasing, and each converges towards some fixed
point by boundedness and continuity.   Furthermore, this system is
globally attractive to $x_0$ if and only if (\ref{discrete system})
is globally attractive to $x_0$.   The conclusion follows
immediately. \epr

Let $I=\R$ or $I=[a,\infty)$ and let $g:I\to I$ be differentiable,
bounded and decreasing.  We say that system (\ref{discrete system})
is \emph{fix point determined} if
\[
\norma{g'(x_0)}\leq 1 \Leftrightarrow \mbox{system (\ref{discrete
system}) is globally attractive towards $x_0$.}
\]
Thus, the global attractiveness of (\ref{discrete system}) is
determined by the slope of $g(x)$ at its unique fix point.  For
instance, it was shown in \cite{Enciso:Sontag:JMB2004} that the
functions $g(x)=A/(K+x)$, $x\geq0$, form fix point determined
systems for every $A,K>0$, since for such functions system
(\ref{discrete system}) is globally attractive and
$\norma{g'(x_0)}<1$;  see also Corollary~\ref{corollary forms}.

An example of a (discontinuous) function which is \emph{not} fix
point determined is

\be{not fpd} g(x)=\left\{ \begin{array}{ll}
1, & x<-0.5, \\
0, & -0.5\leq x \leq 0.5, \\
-1, & x>0.5.
\end{array} \right.
\ee

This function has the unique fix point $x_0=0$ and $g'(0)=0$, but
there is the obvious stable cycle $g(1)=-1,\ g(-1)=1$.  To obtain a
proper example of a differentiable function which is not fix point
determined, it is sufficient to smoothen $g(x)$ above with an
appropriate convolution operator.

The reader will have noticed the importance of $g$ being fix point
determined from the discussion leading to the statement of
Theorem~\ref{main theorem}. Nevertheless $g$ is only defined in
terms of the functions $g_i$, and the composition of fix point
determined functions is not necessarily fix point determined (nor is
the composition of merely sigmoidal functions necessarily
sigmoidal).  This is why the Schwarzian derivative becomes useful at
this point.

\begin{Lemma} \label{lemma determined 1}
 Let $g:I\to I$ be $C^3$, decreasing and bounded, and such that $Sg<0$. Then $g$ is fix point determined.
\end{Lemma}

\bpr Consider the increasing function $G=g^2=g\circ g$, and note
that $G'(x_0)=g'(x_0)^2$. If $\norma{g'(x_0)}>1$, hence $G'(x_0)>1$,
then by boundedness it must follow that $G(z)=z$ for some $z>x_0$.
 Therefore (\ref{discrete system}) has a nontrivial cycle of period 2,
since $g(z)\not=z$.

Conversely, let $G'(x_0)\leq 1$, and assume that $G(z)=z$ for some
$z\not=x_0$.  Without loss of generality we can assume that $G(y)=y,
\ G(z)=z$, for some $y,z$ such that $y<x<z$.  We use here the fact
that $SG<0$, by Lemma~\ref{lemma Sg}. Suppose first that $G(x)=x$ on
some interval containing $x_0$.  Then in the interior of this
interval it would hold $G''(x)=G'''(x)=0$ and thus $SG=0$ for these
points, a contradiction.  It is easy to conclude, using the mean
value theorem, that there exist constants $y_1,z_1$ such that
$y<y_1<x_0<z_1<z$ and $G'(y_1)>1$, $G'(z_1)>1$.  Now consider the
function $G'(x)$ on the interval $[y_1,z_1]$.  The results above
imply that this function has a minimum $w_1$ on the interior of this
interval, and that therefore $G''(w_1)=0,\ G'''(w_1)\geq0$.  Thus
$SG(w_1)\geq0$, a contradiction. \epr

\begin{Corollary} \label{corollary forms} Let $I=\R$ or $I=[a,\infty)$, and let $g:I\to I$
be decreasing and bounded.  If $g(x)$ is the composition of
functions each of which either i) has negative Schwarzian
derivative, or ii) is of Michaelis Menten form for $m\geq 1$, then g
is fix point determined.
\end{Corollary}

\bpr If $g$ is the composition of functions all of which have
negative Schwarzian derivative, then this must be true of $g$ as
well, and $g$ is fix point determined by Lemma~\ref{lemma determined
1}. The same holds if some of the $g_i$ are of Michaelis-Menten form
with $n>1$, by Lemma~\ref{lemma MM}. If some but not all of these
functions are of Michaelis Menten form for $m=1$, then still
$Sg(x)<0$ by the derivation formula in Lemma~\ref{lemma Sg}.

Finally, if all the functions are of the form $(\alpha + \beta
x)/(\gamma + \delta x)$, $\alpha, \beta, \gamma, \delta \geq0$, then
$g$ and $g^2$ are also of this form.  It is then easy to show that
the (bounded, increasing) function $g^2(x)$ is concave down on $I$,
and that it has a unique fixed point $x_0$ which further satisfies
$g'(x_0)^2=(g^2)'(x_0)\leq 1$. The result follows from
Lemma~\ref{lemma equivalence}.  \epr

The relationship between the nonlinear system (\ref{nonlinear
general}) and the discrete system (\ref{discrete system}) becomes
clear in the  proof sketch of the following well-studied result. See
Angeli and Sontag \cite{Angeli:Sontag:interconnections}  and Enciso,
Smith, and Sontag \cite{Enciso:Smith:Sontag:JDE2005} for an abstract
formal treatment, as well as Sontag \cite{Sontag:embedding} for a
discussion of the embedding argument.  The use of the lemma by
Dancer in this context is new.

\begin{Proposition} \label{proposition SGT} Consider a system (\ref{nonlinear general})
under assumption (\ref{assumptions}), and let $g(x)$ be defined by
(\ref{g def}).  If (\ref{discrete system}) is globally attractive
towards $x_0$, then (\ref{nonlinear general}) is globally attractive
towards a unique equilibrium.
\end{Proposition}

\emph{Sketch of Proof:} An elegant result of Dancer
\cite{Dancer:1998} shows that in an abstract monotone system with
bounded solutions and a unique equilibrium, all solutions must
converge towards this equilibrium (the result is stated for discrete
systems in \cite{Dancer:1998}, but a variation for continuous
systems is straightforward). Consider the extended $2n$-dimensional
system

\be{embedding system}
\begin{array}{l}
\dot{x}_i= g_i(x_{i+1}) - \mu_i x_i,\  i=1\ldots n-1, \\
\dot{x}_n= g_n(z_1(t-\tau)) - \mu_n x_n,  \\
\dot{z}_i= g_i(z_{i+1}) - \mu_i z_i,\  i=1\ldots n-1, \\
\dot{z}_n= g_n(x_1(t-\tau)) - \mu_n z_n.  \\
\end{array}
\ee It is not difficult to see that a trajectory $(x_1(t)\ldots
x_n(t))$ is a solution of (\ref{nonlinear general}) if and only if
$(x_1(t)\ldots x_n(t),x_1(t)\ldots x_n(t))$ is a solution of
(\ref{embedding system}).  Moreover, this system is now subject to
positive feedback, since $\delta_1\cdot \ldots \delta_n \cdot
\delta_1\cdot \ldots \cdot \delta_n=1$.  Thus this system is
monotone with respect to a certain partial order; see
\cite{Smith:monotone}, Chapter~5, and \cite{Sontag:multi}. Finally,
the equilibria of this system are in bijective correspondence with
the solutions of $g(g(x))=x$. The conclusion follows by the result
by Dancer and Lemma~\ref{lemma equivalence}.

\vspace{2ex}

\section{Hopf Bifurcation} \label{section Hopf}

In this section we consider the linearization

\be{linear general}
\begin{array}{l}
\dot{x}_i= k_i x_{i+1} - \mu_i x_i,\  i=1\ldots n-1, \\
\dot{x}_n= k_n x_1(t-\tau) - \mu_n x_n,  \\
\end{array}
\ee of system (\ref{nonlinear general}) around its unique
equilibrium point $(\overline{x}_1,\ldots \overline{x}_n)$.  It is
easy to see that

\be{eq ki}
\begin{array}{l}
k_i=g_i'(\overline{x}_{i+1}),\ i=1\ldots n-1, \\
k_n=g_n'(\overline{x}_1).
\end{array}
\ee

We will show in the negative feedback case $k_1\ldots k_n<0$ that
for $\norma{k_1\cdot\ldots \cdot k_n}> \mu_1 \cdot\ldots \cdot
\mu_n$, a Hopf bifurcation exists on the parameter $\tau$. The
characteristic polynomial associated to the linear system
(\ref{linear general}) is \be{polynomial}
g(z,\tau):=(z+\mu_1)(z+\mu_2)\cdot \ldots \cdot (z+\mu_n)+ K
e^{-\tau z}, \ee where $K:=-k_1\cdot\ldots \cdot k_n>0$. See Lemma~3
of Hofbauer and So \cite{So:2000}.

\begin{Lemma} \label{lemma tech 1}
Let $g(\lambda,\tau_0)=0$ for $\lambda=\sigma + \omega i$,
$\tau_0>0$, and assume that $\sigma\geq 0$.  Then there exists an
open neighborhood $U$ of $\tau_0$, and a differentiable function
$\rho:U\to \C$, such that $g(\rho(\tau),\tau)=0$ on $U$. If
$\sigma=0$, then $\mbox{Re }\rho'(\tau_0)>0$.
\end{Lemma}

\bpr Define $f(z):=\prod_i (z+\mu_i)$.  The proof of the first
statement follows by the implicit function theorem for the function
$g(z,\tau)$ at the point $(\lambda,\tau_0)$, after verifying that
$\partial g /\partial z\not=0$ at that point:
\[
\frac{\partial g}{\partial z}(\lambda,\tau_0)= f(\lambda)\sum_j
\frac{1}{\lambda +\mu_j} - \tau_0 K e^{-\lambda \tau_0} =
-Ke^{-\lambda \tau_0} Q(\lambda,\tau_0),\]
where

\[Q(\lambda,\tau_0):=\sum_j \frac{1}{\lambda +\mu_j} + \tau_0.
\]
Using the fact that $\mu_j\geq0$ for every $j$, it is easy to see
that $\mbox{Re }Q(\lambda,\tau_0)>0$ and the proof is complete.

To prove the second statement, let $\sigma=0$.  Note that
necessarily $\omega\not=0$, since $g(z,\tau)>0$ whenever $z\geq0$.
Assume $\omega>0$, the other case being similar.  Multiplying on
both numerator and denominator by $\lambda-\mu_j$, it follows that
\[
\mbox{Im } Q(\lambda,\tau_0)=-\omega \sum_j \frac{1}{\omega^2 +
\mu_j^2} <0.
\]
By the implicit function theorem,

\[
\rho'(\tau_0)=-\frac{\partial g}{\partial \tau}(\lambda, \tau_0)
\left(\frac{\partial g}{\partial \tau}(\lambda, \tau_0)\right)^{-1}
= - \omega i Q(\lambda,\tau_0)^{-1}.
\]
It follows that $\mbox{Re }\rho'(\tau_0)>0$ as stated. \epr

\begin{Theorem} \label{theorem Hopf} If $K>\mu_1\cdot \ldots \cdot \mu_n$, then
system (\ref{nonlinear general}) has a Hopf bifurcation on the
parameter $\tau$.
\end{Theorem}

\bpr

We show that there exists $\tau_0\geq 0$ such that
\begin{description}
\item{i)} $g(\omega i,\tau_0)=0$ for some $\omega>0$,

\item{ii)} $g(\lambda,\tau_0)\not=0$, for all $\lambda \in \C$ with
$\mbox{Re }\lambda >0$,

\item{iii)} for some $\omega_0>0$, it holds
that $g(\omega_0,\tau_0)=0$ and that if $g(\lambda,\tau_0)=0$,
$\lambda=m \omega_0$ for integer $m$ then $\lambda=\pm \omega_0 i$.
\end{description}

Together with Lemma~\ref{lemma tech 1}, this will directly imply the
existence of a Hopf bifurcation at the point $\tau=\tau_0$; see
Theorem~11.1.1 of Hale~\cite{Hale:1993}.

 Let
\[
S:=\{\tau\geq0 \, |\, g(\lambda,\tau)=0 \mbox{ for some $\lambda\in
\C$ such that } \mbox{Re }\lambda\geq0\}.
\]
To see that $S$ is nonempty, first note that whenever $\omega>0$ and
$\norma{f(\omega i)}=K$, one can find $\tau>0$ such that $e^{-\omega
i \tau}=-f(\omega i)/K$ and so $g(\omega i,\tau)=0$.  Noting that
$\norma{f(0)}=\mu_1\cdot \ldots \cdot \mu_n<K$ and $\norma{f(\omega
i)}\to \infty $ as $\omega\to \infty$, it follows by the
intermediate value theorem that $\norma{f(\omega i)}=K$ for some
$\omega$; therefore $S\not=\emptyset$.

Let $\tau_0:=\inf\ S$; it is shown now that $\tau_0\in S$.  Let
$\sigma_1>\sigma_2> \ldots \to \tau_0$, and let
$\lambda_1,\lambda_2, \ldots$ be such that $\mbox{Re }\lambda_i\geq
0$ and $g(\lambda_i,\sigma_i)=0$ for every $i$.  Let $M>0$ be such
that $\norma{f(z)}>2K$ for $\norma{z}\geq M$. Then
$\norma{e^{-\lambda_i \tau}}<1$, and therefore necessarily
$\norma{\lambda_i}<M$, for every $i$.  There exists thus a
subsequence of $\{\lambda_i\}$ which converges towards $\lambda_0\in
\C$, $\mbox{Re }\lambda_0\geq0$.  By continuity
$g(\lambda_0,\tau_0)=0$, and $\tau_0\in S$.

To complete the proof of i) and ii), it suffices to show that
$g(\lambda,\tau_0)=0$, $\mbox{Re }\lambda \geq0$ imply $\mbox{Re
}\lambda=0$. But this follows directly from Lemma~\ref{lemma tech
1}, by the minimality of $\tau_0$.

To see iii), simply recall that $g(\omega i,\tau_0)=0$ implies
$\norma{\omega i}<M$, and pick $\omega_0>0$ so that $\omega_0 i$ is
a root with maximal norm.
\epr

Note that this result is proved in the context of Theorem~11.1.1 of
\cite{Hale:1993}.  The existence of periodic solutions for certain
values $\tau>\tau_0$ follows, but no assertion is made regarding
their stability.  This may nevertheless be shown using the above
proof, if the asymptotic stability of the equilibrium of
(\ref{nonlinear general}) is established for $\tau=\tau_0$.

In the particular case $\tau=0$, it is known \cite{Tyson:1978} that
system (\ref{linear general}) is asymptotically stable provided that
$K/(\mu_1\cdot \ldots \cdot \mu_n)< \sec^n(\pi/n)=1/(cos^n(\pi/n))$.
Therefore necessarily $\tau_0>0$ in those cases.

The following proposition establishes a global stability result for
the linear system (\ref{linear general}).

\begin{Proposition}
Let $K>\mu_1\cdot \ldots \cdot \mu_n$.  Let $\tau_0\geq0$ be the
Hopf bifurcation point as in Theorem~\ref{theorem Hopf}, and let
$\tau\geq0$. Then the linear system (\ref{linear general}) is
exponentially unstable if and only if $\tau>\tau_0$.
\end{Proposition}

\bpr It was shown in the proof of Theorem~\ref{theorem Hopf} that if
$\tau<\tau_0$ ($\tau=\tau_0$), then $g(\cdot,\tau)$ could have no
root $\lambda$ with $\mbox{Re } \lambda \geq0$ ($\mbox{Re } \lambda
> 0$).  Therefore for $\tau\leq \tau_0$, the exponential stability
of (\ref{linear general}) is ruled out.

Let $S'$ be the set of $\tau\geq0$ such that system (\ref{linear
general}) is exponentially unstable.  It follows from
Lemma~\ref{lemma tech 1} and the proof of Theorem~\ref{theorem Hopf}
that $(\tau_0,\tau_0+\epsilon)\subseteq S'$ for some $\epsilon>0$.
Assume by contradiction that $S'\not=(\tau_0,\infty)$, and let
$\tau_1$ be the infimum of $(\tau_0,\infty)-S'$.  In particular, it
holds that $\tau_1\geq \tau_0 +\epsilon>\tau_0$.  But this is once
again a violation of Lemma~\ref{lemma tech 1}. \epr

\subsection{Proof of Theorem~\ref{main theorem}}

Now the proof of the main result is complete.

\bpr Let $x_0$ be the unique fix point of $g(x)$. The first case
corresponds to the situation in which $\norma{g'(x_0)}\leq 1$. Since
(\ref{discrete system}) is fix point determined by
Corollary~\ref{corollary forms} and Lemma~\ref{lemma determined 1},
it holds that (\ref{discrete system}) is globally attractive to
equilibrium.  By Proposition~\ref{proposition SGT}, system
(\ref{nonlinear general}) is also globally attractive towards a
unique equilibrium.

In the case $\norma{g'(x_0)}>1$, system (\ref{discrete system}) must
have a periodic solution since it is fix point determined.
Evaluating $g'(x)$ using the chain rule yields that $K>\mu_1\cdot
\ldots \cdot \mu_n$, and therefore by Theorem~\ref{theorem Hopf}, a
Hopf bifurcation occurs on the parameter $\tau$. \epr

\section{The Positive Feedback Case} \label{section positive}

We state and prove the corresponding statement in the positive
feedback case, which follows from the material in \cite{Sontag:mono}
and \cite{Enciso:Sontag:setchars}. Given a system (\ref{nonlinear
general}), consider the positive feedback hypotheses

\be{positive assumptions} \mu_i>0,\ \delta_i g_i(x) \mbox{ strictly
increasing, } \delta_i\in \{1,-1\}, \ i=1\ldots n, \mbox{ and }
\delta_1\cdot\ldots \cdot \delta_n=1. \ee

It is easy to see, by setting the RHS of (\ref{nonlinear general})
equal to zero, that the function $x \in \R \to \phi(x)=(x_1,\ldots
x_n)$ defined by $x_n:=\mu_n^{-1}g_n(x)$,
$x_{n-1}:=\mu_{n-1}^{-1}g_{n-1}(x_n)$, \ldots,\
$x_1:=\mu_1^{-1}g_1(x_2)$ is a bijection between equilibria of
(\ref{discrete system}) and equilibria of (\ref{nonlinear general}).
In the following result, `almost every' is meant in the sense of
measure, more precisely in the sense of prevalence in
\cite{Yorke:1992}.  For other notation used in the proof, refer to
\cite{Enciso:Sontag:setchars}.

\begin{Theorem}
Consider the system (\ref{nonlinear general}) under (\ref{positive
assumptions}), and let $g$ be bounded and have countable equilibria.
Then almost every solution of (\ref{discrete system}) converges
towards some equilibrium $x$ such that $g'(x)\leq 1$. Also, almost
every solution of (\ref{nonlinear general}) converges towards some
equilibrium $\phi(x)$ such that $g'(x)\leq 1$.
\end{Theorem}

\bpr The first part of this theorem is straightforward: every
solution of (\ref{discrete system}) is monotone increasing or
decreasing, and it is bounded since $g$ is bounded.  Therefore each
solution must converge towards an equilibrium.  But if $x$ is an
equilibrium of (\ref{discrete system}) such that $g'(x)>1$, then it
is repelling and no strictly monotone solution can converge towards
$x$.  Since $g$ is strictly increasing and therefore injective,
given any $z\in I,\ k\geq1$ it holds that $g^k(z)=x$ implies $z=x$.
Thus the only solution that converges towards $x$ is that with
initial condition $x$.

The second statement follows from Theorem~6 of
\cite{Enciso:Sontag:setchars}, which is a consequence of results in
\cite{Sontag:mono,Hirsch:1988}.  It follows from the results in
\cite{Enciso:Sontag:DCDS2006}, or from Chapter~5 of
\cite{Smith:monotone} after a change of variables, that
(\ref{nonlinear general}) is monotone with respect to an orthant
cone.  The strong monotonicity follows using the strict monotonicity
of the functions $g_i(x)$ in (\ref{positive assumptions}); see the
proof of Theorem~1.3 in \cite{Smith:Hirsch:Posta2003}. The
boundedness of $g$ implies that one of the functions $g_i$ is
bounded, and therefore all solutions of the system are also bounded.

We write system (\ref{nonlinear general}) as the closed loop of a
control monotone system by replacing $g_n(x_1(t-\tau))$ on the right
hand side by $g_n(u)$ and by letting $h(x_t)=x_1(t-\tau)$. Theorem~6
of \cite{Enciso:Sontag:setchars} implies in this case that almost
all solutions converge towards equilibrium points $e=(x_1,\ldots
x_n)=\phi(x)$ such that either the linearization (\ref{linear
general}) is not irreducible, $g_n'(x_1)=0$, or else $g'(x)\leq 1$.
But any of the first two conditions imply that in the linearization
(\ref{linear general}) it holds $k_i=0$ for some $i$, and that
therefore

\[g'(x)=k_1\cdot \ldots k_n/(\mu_1\cdot \ldots \cdot \mu_n)=0\leq 1.\]
Thus almost every
solution converges to an equilibrium $e=\phi(x)$ such that
$g'(x)\leq1$. \epr

\section{Future Work} \label{section future}

The framework of Angeli and Sontag \cite{Sontag:mono} and Enciso,
Smith and Sontag \cite{Enciso:Smith:Sontag:JDE2005} describes quite
general dynamical systems as the negative feedback loop of
controlled monotone systems.  Sufficient conditions are then given
for the system to be globally attractive to equilibrium, even in the
presence of delays or diffusion terms.   Theorem~\ref{main theorem}
can potentially be used to extend these results to the case of
periodic oscillations, as well as to show that the original results
are sharp in some sense.  It is not the first time that this is
suggested.  For instance, Angeli and Sontag
\cite{Angeli:Sontag:interconnections} have pointed out that if the
associated discrete system has a $2$-cycle, then large enough delays
would create the appearance of oscillatory behavior (or
\emph{pseudooscillations}), which in a biological system might be as
meaningful as proper periodic oscillations.

The analysis of the asymptotic behavior of the system considered in
this paper is far from complete.  If the system falls into the
second case of the main theorem, simulations suggest that for $\tau
> \tau_0$ the system is in fact globally attractive towards a unique
nonconstant periodic solution. Work towards such a result would most
likely include the use of the Poincare-Bendixson result, for example
by finding a Morse decomposition of the system and ruling out the
existence of homoclinic orbits.

Finally, note that the need for the assumption $Sg<0$ can be traced
back to the particular approach used to prove the existence of
periodic oscillations (Hopf bifurcation), in the following sense: if
one could prove the existence of periodic oscillations of
(\ref{nonlinear general}) based solely on the existence of a stable
periodic 2-cycle of (\ref{discrete system}), then the proof of the
main theorem wouldn't have to require that $g$ is fix point
determined, and the assumption $Sg<0$ could be dropped.  Indeed, it
has been observed in numerical simulations that whenever there is a
stable 2-cycle of (\ref{discrete system}), then there is also a
limit cycle of (\ref{nonlinear general}) for large enough $\tau$ ---
\emph{even when $Sg\not< 0$}. This has been numerically observed to
be also true in more complex noncyclic systems in the framework of
\cite{Enciso:Smith:Sontag:JDE2005}.   Note that the proof of the
existence of periodic solutions would require to abandon any obvious
use of Hopf bifurcation or ejective fixed point methods, since it
could not be required that $\norma{g'(x_0)}>1$.

\paragraph{Acknowledgments:} The author would like to thank his
former advisor Eduardo Sontag for many helpful discussions and
comments.  Also to be thanked are David Angeli, Tomas Gedeon,
Benjamin Kennedy and Roger Nussbaum, for their friendly input and
correspondence.


\end{document}